\documentclass[conference]{IEEEtran}
\ifCLASSINFOpdf
\else
\fi
\usepackage{amsmath,dsfont,bbm,epsfig,amssymb,amsfonts,amstext,verbatim,amsopn}
\usepackage{cite,subfigure,multirow,multicol,lipsum,xfrac}
\usepackage{amsthm}
\usepackage{mathtools,amsthm}
\usepackage{perpage}
\usepackage{balance}
\usepackage{url}
\usepackage{amsfonts}
\usepackage{epsfig}
\usepackage[font={small}]{caption}
\usepackage{psfrag}	
\usepackage{etoolbox}
\usepackage{algorithmicx}
\usepackage[Algorithm,ruled]{algorithm}
\usepackage{algpseudocode}
\usepackage{pifont}
\usepackage[utf8]{inputenc}
\usepackage[T1]{fontenc}  
\usepackage[nolist]{acronym}
\MakePerPage{footnote}
\usepackage{paralist}
\usepackage{enumitem}
\usepackage{bbm}
\usepackage[process=auto]{pstool}
\usepackage{tikz}
\usetikzlibrary{shapes,arrows}

\newcommand{\setR}{\mathbbmss{R}}

\newcommand{\setS}{\mathbbmss{S}}

\newcommand{\setZ}{\mathbbmss{Z}}

\newcommand{\setC}{\mathbbmss{C}}

\newcommand{\rmF}{\mathrm{F}}

\newcommand{\her}{\mathsf{H}}

\newcommand{\sfM}{\mathsf{M}}

\newcommand{\m}{{\mathsf{max}}}

\newcommand{\mao}{\mathcal{O}}

\newcommand{\mai}{\mathcal{I}}

\newcommand{\bhh}{\mathbf{h}}

\newcommand{\bx}{{\boldsymbol{x}}}

\newcommand{\set}[1]{\left\lbrace#1\right\rbrace}

\newcommand{\dif}{\mathrm{d}}

\newcommand{\by}{{\boldsymbol{y}}}

\newcommand{\bn}{{\boldsymbol{n}}}

\newcommand{\mI}{\mathbf{I}}

\newcommand{\mX}{\mathbf{X}}

\newcommand{\mH}{\mathbf{H}}

\newcommand{\loge}{\mathrm{ln}\hspace*{.5mm}}

\newcommand{\E}{\mathsf{E}\hspace{.5mm}}

\newcommand{\norm}[1]{\lVert #1 \rVert}

\newcommand{\abs}[1]{\lvert #1 \rvert}
\newcommand{\tr}[1]{\mathrm{Tr} \{ #1 \}}

\newtheoremstyle{mystyle}%                % Name
  {}%                                     % Space above
  {}%                                     % Space below
  {}%                                     % Body font
  {}%                                     % Indent amount
  {\bfseries}%                            % Theorem head font
  {:}%                                     % Punctuation after theorem head
  { }%                                    % Space after theorem head, ' ', or \newline
  {}%                                     % Theorem head spec (can be left empty, meaning `normal')

\theoremstyle{mystyle}

\newtheorem{definition}{Definition}

\newtheorem*{prf}{Proof}

\newtheorem{theorem}{Theorem}
\newcounter{bar}

\begin{document}
\title{Asymptotic Performance Analysis of Spatially Reconfigurable Antenna Arrays}

% Authors
\author{
\IEEEauthorblockN{
Saba Asaad\IEEEauthorrefmark{2}\IEEEauthorrefmark{1},
Ali Bereyhi\IEEEauthorrefmark{1},
Mohammad Ali Sedaghat\IEEEauthorrefmark{1},
Ralf R. M\"uller\IEEEauthorrefmark{1},
Amir M. Rabiei\IEEEauthorrefmark{2}
}
\IEEEauthorblockA{
\IEEEauthorrefmark{2}School of Electrical and Computer Engineering, University of Tehran\\
\IEEEauthorrefmark{1}Institute for Digital Communications (IDC), Friedrich-Alexander Universit\"at Erlangen-N\"urnberg (FAU)\\
saba{\_}asaad@ut.ac.ir, ali.bereyhi@fau.de, mohammad.sedaghat@fau.de, ralf.r.mueller@fau.de, rabiei@ut.ac.ir
\thanks{This work was supported by the German Research Foundation, Deutsche Forschungsgemeinschaft (DFG), under Grant No. MU 3735/2-1.}
}
}


\IEEEoverridecommandlockouts

\maketitle

\begin{abstract}
A spatially reconfigurable antenna arrays consists of an antenna array of finite length and fixed geometry which is displaced within a given area. Using these reconfigurable components, the performance of MIMO systems is remarkably improved by effectively positioning the array in its displacement area. This paper studies the large-system performance of MIMO setups with spatially reconfigurable antenna arrays when the displacement area is large. Considering fading channels, the distribution of the input-output mutual information is derived, and the asymptotic hardening property is demonstrated to hold. As the size of the displacement area grows large, the mutual information is shown to converge in distribution to a type-one Gumbel random variable whose mean grows large proportional to the displacement size, and whose variance tends to zero. Our numerical investigations depict that the type-one Gumbel approximation closely tracks the empirical distribution even for a finite displacement size.\\
%a perfect match of the empirical and converging type-one Gumbel distribution for a finite size of the displacement area. \\

\end{abstract}


\IEEEpeerreviewmaketitle

\section{Introduction}
\label{sec:intro}
\ac{mimo} systems have been shown to enhance the throughput significantly. The \ac{mmw} spectrum enables us to employ large antenna arrays at the both user terminal and access point~\cite{rappaport2013millimeter}.~Consequently, massive \ac{mimo} systems have been considered as a key technology for the next generation of mobile communications \cite{marzetta2010noncooperative}. Despite the promising results obtained through theoretical investigations, these systems have still remained challenging from the implementational point of view. Consequently, several ideas such as antenna selection \cite{molisch2005capacity} have been addressed in the context of massive \ac{mimo} systems in the \ac{mmw} spectrum, in order to reduce the overall \ac{rf} cost of the system \cite{li2014energy,asaad2017tas}. These ideas mainly propose solutions based on the reduction in the number of \ac{rf} chains while keeping the array size large. Spatially reconfigurable antenna arrays \cite{gheethan2015passive} suggest an alternative solution by employing a finite number of antennas displacing within a given area. In \ac{mmw}, these reconfigurable arrays can acquire substantial performance improvements due to the fact that a displacement area of practical sizes is relatively large compared to physically small antenna arrays. The asymptotic performance of these configurations, however, have not yet been addressed in the literature precisely.

\subsection*{Spatially Reconfigurable Antenna Arrays}
In spatially reconfigurable antenna arrays, an antenna array of finite length and fixed geometry is being displaced through a given area. This setup allows for effectively changing the array's location at the beginning of each transmission interval such that the best channel is observed, and consequently, leads to remarkable performance improvements. As the displacement area of a spatially reconfigurable antenna array grows large, efficient positioning methods demonstrate the same limiting behavior as massive \ac{mimo} systems in the \ac{mmw} spectrum. These properties are obtained at the expenses of a large displacment area and challenges raising in the~design~of~reconfigurable arrays. However, compared to the high \ac{rf} cost, the issues can be more effiectively coped with in practice. In fact, these issues can be resolved by moving to the mmWave spectrum, since a large displacement area is realized within a rather small physical platform \cite{rappaport2013millimeter}. Recent proposals for the implementation of these arrays, such as microfluidically reconfigurable arrays \cite{gheethan2015passive,dey2016microfluidically,palomo2016microfluidically,yilmaz2016millimeter}, moreover, have efficiently addressed the latter issue.

\subsection*{Asymptotic Hardening Property}
Consider the Gaussian \ac{mimo} channel $\mH_{N_r\times N_t}$ with ${N_t}$ tr- ansmit and ${N_r}$ receive antennas, and assume the \ac{snr} at each receive antenna is $\rho$. In this case, the input-output mutual information, when \ac{iid} zero-mean unit-variance Gaussian symbols are transmitted through the channel, is given by
\begin{align}
\mai = \log { \det\abs{ \mI_{N_r}+ \rho \mH \mH^\her  }}.
\end{align}
Let $\mH$ be an \ac{iid} unit-variance Rayleigh fading channel. As $N_t\uparrow\infty$, for a fixed $N_r$, $N_t^{-1}\mH\mH^\her$ converges to the covariance matrix of the column vectors, i.e., $\mI_{N_r}$, and therefore, the mutual information converges to
\begin{align}
\mai = N_r \log \set{ { 1+ \rho N_t  }}.
\end{align}
Consequently, one can conclude that the channel converges to a deterministic channel, and the effect of fading vanishes as $N_t$ grows large. This asymptotic property of the channel is called the ``hardening property'' and was initially studied~in~\cite{hochwald2004multiple} where the authors showed that the mutual information is normally distributed around its mean while the variance tends rapidly to zero. The property was further shown to hold under the \ac{tas} when a finite number of active antennas are selected from a large set of transmit antennas \cite{bai2007channel,li2014energy,asaad2017tas}.

%was first studied in [7]. The property indicates that in the
%large system limit, the distribution of the mutual information
%between a white Gaussian input and the output of a MIMO
%Gaussian fading channel concentrates almost normally around
%its mean while the variance shrinks rapidly as the dimension
%grows large. Under single \ac{tas} (TAS),
%the channel hardening was studied initially in [4], where the
%authors considered a TAS protocol selecting a single transmit
%antenna with the strongest channel gain. The authors further
%showed that, under this scheme, the channel hardens at a
%slower rate compared to the case considered in [7]. In [5], the
%limiting behavior of mutual information in an uplink channel
%was investigated considering the transmitter to be equipped
%with a single transmit antenna, and the receiver to select a
%certain number of strongest channels. For this scenario, the
%distribution of input-output mutual information was approximated
%with the distribution of the logarithm of a folded normal
%random variable. In the large system limit, it was further
%shown that the variance converges to zero which concluded the
%asymptotic hardening property for the setup. The asymptotics
%of TAS were further studied in [6]

\subsection*{Contributions}
In this paper, we study spatially reconfigurable transmit antenna arrays with large displacement areas when the array is positioned such that the strongest channel between the transmit array and receive antennas is observed. To investigate the performance in the large limit, we model a spatially reconfigurable array by introducing the concept of virtual channel. The positioning of the array therefore reduces to a \ac{tas} problem in which the set of strongest neighboring antennas are selected from a set of large transmit antennas. For this setup, we deter-mine the distribution of the input-output mutual information and show that, as the size of displacement area grows large, the mutual information converges in distribution to a type-one Gumbel random variable whose mean grows large, and whose variance converges to zero; the result which demonstrate the asymptotic hardening property of large spatially reconfigurable antenna arrays.

\subsection{Notation}
We represent scalars, vectors and matrices with non-bold, bold lower case and bold upper case letters, respectively. A $K \times K$ identity matrix is shown by $\mI_K$, and $\mH^{\her}$ indicates the Hermitian of the matrix $\mH$. The set of real and integer numbers are denoted by $\setR$ and $\setZ$, and their corresponding non-negative subsets are shown by superscript $+$; moreover, $\setC$ represents the complex plane. The cardinality of the set $\setS$ is indicated by $\abs{\setS}$. $\norm{\bx}$ denotes the Euclidean norm of the vector $\bx$. For a random variable $x$, $\mathrm{f}_x$ and $\rmF_x$ represent the \ac{pdf} and \ac{cdf}, respectively. $\log\set{\cdot}$ and $\loge \set{\cdot}$ indicate the binary and natural logarithm, and $\E\set{\cdot}$ is the expectation operator. A type-one Gumbel distribution, or double exponential distribution, with the location factor $\mu$ and scale factor $\sigma$ is denoted by $\mathrm{Gumbel}\hspace*{.5mm} [\mu;\sigma]$ which reads
\begin{align}
\rmF_z(z)= e^{-e^{-\frac{z-\mu}{\sigma}}}
\end{align}
for some random variable $z\sim\mathrm{Gumbel}\hspace*{.5mm} [\mu;\sigma]$.

\section{System Model}
\label{sec:sys}
Consider a single-user Gaussian \ac{mimo} system in which the transmitter is equipped with a spatially reconfigurable antenna array of length $N_t$, and the user has $N_r$ receive antennas. We consider a block fading channel model in which the \ac{csi} is assumed to be constant within a coherence time interval. At the beginning of each time interval, the transmitter selects the location of the antenna array such that the array observes the strongest channels to the user. Our goal is to investigate input-output mutual information for this setup when the number of possible array locations grows large. We tackle this objective by first introducing a selection based model for spatially reconfigurable antenna arrays, and then studying the proposed model.

\subsection{Channel Model}
\label{sec:channel}
In spatially reconfigurable antenna arrays, the antenna array has a degree of freedom to displace within a given area. Let the set $\setS\subseteq \setR^3$, being referred to as the ``displacement support'', denote the area in which the the antenna array can displace. In this case, the received vector $\by_{N_r \times 1}$, at each time interval, is written as
\begin{align}
\by=\sqrt{\rho} \hspace*{1.1mm}\mH\left(x,y,z\right) \hspace*{.1mm} \bx + \bn \label{eq:1}
\end{align}
where $\rho$ denotes the average \ac{snr} at each receive antenna, $\bn_{N_r \times 1}$ is a circularly symmetric zero-mean complex Gaussian noise vector with unit variance, i.e., $\bn \sim \mathcal{CN} (\boldsymbol{0}, \mI_{N_r})$, $\bx_{N_t \times 1}$ identifies the transmit signal with the power constraint $\E \bx^\her \bx \leq {N_t}$, the tuple $(x,y,z) \in \setS$ denotes the coordinate of the array's location, and $\mH(x,y,z) \in \setC^{N_r \times N_t}$ represents the channel between the transmit antenna array at $(x,y,z)$ and the receive antennas. The statistics of the channel $\mH(x,y,z)$ depend on the operating carrier frequency and antennas geometry at the transmit and receive arrays and can be modeled by known channel models, e.g. Kronecker model. The channels at different locations, moreover, might be correlated, since the corresponding locations can be in a relatively small distance compared to the~array~size.% More precisely, assuming the coordinates $(x_1,y_1,z_1)$ and $(x_2,y_2,z_2)$ to be taken from the displacement support,
%\begin{align}
%\E \tr{\mH(x_1,y_1,z_1)^\her \mH(x_2,y_2,z_2)}\neq 0.
%\end{align}
%We assume the transmit antennas in the array, as well as the receive antennas, are spaced enough.

\subsection{General Spatially Reconfigurable Antenna Arrays}
\label{sec:array}
\begin{figure}[t]
%\hspace*{-1.3cm}  
\centering
\resizebox{1\linewidth}{!}{
\pstool[width=.35\linewidth]{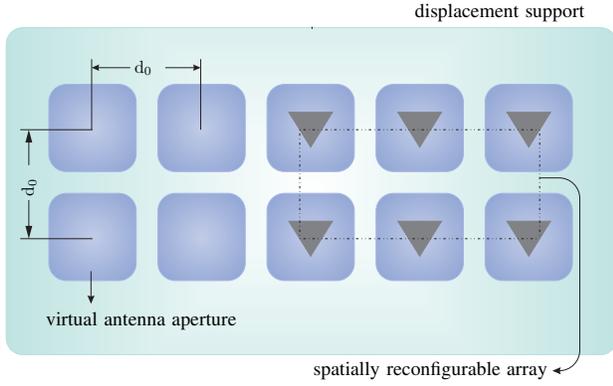}{

\psfrag{displacement support}[l][l][0.28]{displacement support}
\psfrag{virtual antenna location}[l][l][0.28]{virtual antenna aperture}
\psfrag{spatially adaptive array}[l][l][0.28]{spatially reconfigurable array}
\psfrag{1.25}[l][l][0.25]{$\dif_0$}
%%y-axis
%
%\psfrag{-5}[r][c][0.18]{$-5$}
%\psfrag{-10}[r][c][0.18]{$-10$}
%\psfrag{-15}[r][c][0.18]{$-15$}
%\psfrag{-20}[r][c][0.18]{$-20$}
%\psfrag{-25}[r][c][0.18]{$-25$}
%\psfrag{-30}[r][c][0.18]{$-30$}
%\psfrag{0}[r][c][0.18]{$0$}
%%
%%%x-axis
%\psfrag{1}[c][b][0.18]{$1$}
%\psfrag{1.2}[c][b][0.18]{$1.2$}
%\psfrag{1.4}[c][b][0.18]{$1.4$}
%\psfrag{1.6}[c][b][0.18]{$1.6$}
%\psfrag{1.8}[c][b][0.18]{$1.8$}
%\psfrag{2}[c][b][0.18]{$2$}
%\psfrag{2.2}[c][b][0.18]{$2.2$}
%\psfrag{2.4}[c][b][0.18]{$2.4$}
%\psfrag{2.6}[c][b][0.18]{$2.6$}
%\psfrag{2.8}[c][b][0.18]{$2.8$}
}}
\caption{A spatially reconfigurable antenna array modeled by a virtual channel. The imaginary transmit antennas are located at the center of the blue apertures. At each location, the transmit array selects a subset of the apertures.}
\label{fig:1}
\end{figure}

The key point for describing the spatially reconfigurable antenna arrays is to address the correlation among the observed channels. We model the correlation by introducing the concept of the ``virtual channel'' over $\setS$. To illustrate the idea, consider two given locations at $(x_1,y_1,z_1)$ and $(x_2,y_2,z_2)$ with the Euclidean distance $\dif$, and denote their corresponding observed channels with $\mH_1$ and $\mH_2$, respectively, i.e.,
\begin{subequations}
\begin{align}
\mH_1&=\mH\left(x_1,y_1,z_1\right),\\
\mH_2&=\mH\left(x_2,y_2,z_2\right).
\end{align}
\end{subequations}
For large $\dif$, $\mH_1$ and $\mH_2$ can be uncorrelated. However, as the distance reduces, the channels become strongly correlated, and consequently indistinguishable. Therefore, we define the effective step size $\dif_0$ which is considered to be the same as the distance between two neighboring antennas in the transmit array, and assume the transmit array to displace by a multiple of $\dif_0$. Based on this assumption, one can grid $\setS$ into a finite set of location points in which the space between each two neighboring locations is $\dif_0$. Let us denote the number of locations by $L$ and their corresponding coordinates by $(x_\ell,y_\ell,z_\ell)$ for all $\ell \in [1:L]$. In this case, one can think of $L$ imaginary transmit antennas located at $(x_\ell,y_\ell,z_\ell)$, and therefore, the displacement support is represented by the channel between these imaginary antennas and the receiver. We call this channel, the virtual channel over the displacement support $\setS$, and denote it by the $N_r\times L$ matrix $\mH(\setS)$. Considering the virtual channel, the positioning of the spatially reconfigurable antenna array is then modeled as the selection of $N_t$ neighboring transmit antennas over $\mH(\setS)$. Fig.~\ref{fig:1} illustrates the model based on the virtual channel for a spatially reconfigurable antenna array. As it is shown, the displacement support is divided into a set of apertures which are spaced from each other by distance $\dif_0$. By considering an imaginary transmit antenna at the center of each aperture, the specially reconfigurable antenna array virtually selects $N_t$ neighboring apertures within the displacement support.

The channel between the transmit array at $(x_\ell,y_\ell,z_\ell)$ and the receive antennas is determined in terms of the virtual channel $\mH(\setS)$. To illustrate the latter statement, let $M$ be the number of all possible subsets of $N_t$ neighboring apertures. For a given $m\in [1:M]$, denote the set of the corresponding aperture indices with $\setS_m$ where $\abs{\setS_m}=N_t$ and $\setS_m \subset [1:L]$. In this case, by assuming that the transmit array at $(x_\ell,y_\ell,z_\ell)$ is located in the $m$th subset of neighboring apertures, $\by$ in \eqref{eq:1} is rewritten as
\begin{align}
\by= \sqrt{\rho} \hspace*{.6mm} \mH_m \hspace*{.5mm} \bx + \bn
\end{align}
where $\mH_m$ is an $N_r \times N_t$ matrix constructed from $\mH(\setS)$ by collecting the columns which correspond $\setS_m$.

\subsection{One-Dimensional Spatially Reconfigurable Antenna Arrays}
Due to hardware limitations, the spatially reconfigurable antenna arrays are usually considered to displace in~one~dimension. This approach fits new hardware proposals such as microfluidically reconfigurable \ac{rf} devices \cite{gheethan2015passive,dey2016microfluidically,palomo2016microfluidically, yilmaz2016millimeter}.~Considering a one-dimensional spatially reconfigurable array, the number of possible subsets of neighboring apertures reads $M=L-N_t+1$, and the columns of $\mH_m$, for any index $m$, are neighboring columns in $\mH(\setS)$ which means that
\begin{align}
\setS_m=[m:m+N_t-1]
\end{align}
for $m\in[1:L-N_t+1 ]$. Therefore, by denoting $\mH(\setS)$ as
\begin{align}
\mH(\setS)=\left[ \bhh_1, \ldots, \bhh_L \right],
\end{align}
with $\bhh_\ell$ being an $N_r\times1$ vector, $\mH_m$ is given by
\begin{align}
\mH_m=\left[ \bhh_m, \ldots, \bhh_{m+N_t-1} \right]. \label{eq:Hm}
\end{align}
In the sequence, we focus on~one-dimensional spatially reconfigurable antenna arrays.

\section{Problem Formulation}
\label{sec:prob}
Considering the model proposed in Section \ref{sec:sys}, we investigate the performance of a one-dimensional spatially reconfigurable antenna array considering a single receive antenna, i.e., $N_r=1$, and the virtual channel $\mH(\setS)$ to be an \ac{iid} Rayleigh fading channel. The former assumption is common in downlink scenario of cellular networks, and the latter fits the model given in Section \ref{sec:sys} at the frequencies of which $\dif_0$ is larger than the decorrelation distance defined as the minimum distance between two antennas in which their received symbols are approximately uncorrelated. We leave the extension of the analysis to cases with multiple receive antennas and correlated channel models as possible future works. Although the virtual channel $\mH(\setS)$ and observed channels $\mH_m$ reduce to row vectors as $N_r=1$, we do not change their notation for sake of compactness.% Consequently, the sequence of observed channels are modeled by an $N_t$-dependent sequence.
\begin{definition}[\bfseries $\sfM$-Dependent Sequences]
\label{def:k-dep}
A sequence of random objects $\mX_1, \ldots, \mX_n$ is said to be $\sfM$-dependent, if for any two indices $i$ and $j$ such that $\abs{i-j} \geq \sfM$, $\mX_i$ and $\mX_j$ are independent.
\end{definition}
Regarding \eqref{eq:Hm}, the sequence of the observed channels is an $N_t$-dependent sequence, since $\mH_m$ overlaps only with $N_t-1$ next or previous selected channels.

\subsection{Array Positioning}
\label{ssec:local}
At the beginning of each transmission interval, a pilot signal is transmitted while the transmit antenna array is displaced through the whole displacement support. The \ac{csi}, namely~the~observed channels $\mH_m$, is therefore supposed to be available at the receiver side. We moreover assume that the transmitter is aware of the location of the antenna array in which the observed channel has the most dominant or strongest coefficients. This information is provided either by 
\begin{enumerate}[label=(\alph*)]
\item assuming the transmitter estimating the observed channels itself, or
\item assuming the index which corresponds to the dominant observed channel is given to the transmitter through a rate-limited return channel.
\end{enumerate}
In either case, the transmitter locates the antenna array at the corresponding location. In this case, the observed channel is denoted by $\mH_{m^*}$ which reads 
\begin{align}
{m^*} = \arg \max_{m} \tr{\mH_m \mH_m^\her}
\end{align}
for $m\in[1:L-N_t+1]$. 

For this positioning protocol, our objective is to investigate the distribution of the input-output mutual information of the system when the displacement support of the spatially reconfigurable antenna array grows large relative to the distance $\dif_0$, i.e., large $L$.

\section{Main Results}
\label{se:result}
We assume that the antennas transmit \ac{iid} complex Gaussian symbols with equal power. In this case, due to the average transmit power constraint, $\bx\sim\mathcal{CN} (\boldsymbol{0}, N_t^{-1} \mI_{N_t})$. The~input-output~mutual information, therefore, reads
\begin{align}
\mai \coloneqq \mathrm{I} \left( \bx;\by|\mH(\setS) \right)= \log \left( 1 + {\rho} \norm{\mH_{m^*}}^2 \right). \label{eq:mai}
\end{align}
In case that the \ac{csi} is available only at the receiver side, the input-output mutual information in \eqref{eq:mai} represents the maximum achievable rate over the observed channel $\mH_{m^*}$.~However, when the \ac{csi} is available at the both sides, $\mai$ determines only an achievable rate which is a lower bound on the capacity.

The mutual information in \eqref{eq:mai} is in general a random variable whose distribution depends on the distribution~of~$\mH_{m^*}$. Our main result shows that, as $L$ grows large, the distribution of $\mai$ converges to a type-one Gumbel distribution whose mean grows large proportional to $L$.

\begin{theorem}
\label{thm:1}
Consider the one-dimensional transmit antenna array with positioning protocol proposed in \ref{ssec:local}.~As~$L$~grows large, the input-output mutual information $\mai$ converges~in~distribution to the type-one Gumbel random variable $\mai_{\mathrm{asy}}$
\begin{align}
\mai_{\mathrm{asy}}\sim \mathrm{Gumbel} \left[ \mu_L - \gamma \sigma_L ; \sigma_L \right]
\end{align}
where $\gamma$ denotes Euler's constant, and $\mu_L$ and $\sigma_L$ are given~by
\begin{subequations}
\begin{align}
\mu_L &=\log \set{ 1+{\rho} \hspace*{.5mm} \Theta_L } \label{eq:9a} \\
\sigma_L &= \frac{ \beta_L \hspace*{.5mm} \rho \log e}{1+\rho \hspace*{.5mm} \Theta_L } + \mao \left( \frac{1}{\loge^2 L} \right). \label{eq:9b}
\end{align}
\end{subequations}
In \eqref{eq:9a} and \eqref{eq:9b}, $\Theta_L = \xi_L+\phi_L + \gamma \beta_L$ where $\beta_L$ is determined in terms of $\phi_L$ and $\xi_L$ as
\begin{align}
%\hspace*{-2mm}\Theta_L \hspace*{-.7mm}=\hspace*{-.7mm} \xi_L \hspace*{-.7mm}+\hspace*{-.7mm} \phi_L \hspace*{-.7mm}+
\beta_L = \frac{ \left( \xi_L+\phi_L \right)^2+ \left( N_t-1 \right) \left( \xi_L+\phi_L \right)}{\left( \xi_L+\phi_L \right)^2-(N_t-1)(N_t-2)},
\end{align}
$\phi_L$ is given in terms of $\xi_L$ by
\begin{align}
\phi_L \hspace*{-.7mm} = \hspace*{-.7mm} \delta_{N_t,1} \frac{N_t-1}{\xi_L} \left[ 1+ \left( N_t+\xi_L-1 \right) \hspace*{.5mm} \loge \set{\frac{\xi_L}{N_t-1}} \right] %\nonumber
\end{align}
and $\xi_L$ is defined in terms of $L$ and $N_t$ as
\begin{align}
\xi_L= \loge \set{\frac{L-N_t +1}{(N_t-1)!}} +\delta_{N_t,1} (N_t-1) \hspace*{.5mm} \loge(N_t-1)
\end{align}
for $\delta_{N_t,1}$ being equal to zero for $N_t=1$ and one elsewhere.
\end{theorem}
%\Xi_L=\hspace*{1mm}&\loge\frac{L}{(N_t-1)!} \nonumber \\
%& + \frac{N_t-1}{\xi_L} \left[ 1+ \xi_L \hspace*{.5mm} \loge \xi_L + (N_t-1) \hspace*{.5mm} \loge \frac{\xi_L}{N_t-1} \right]
\begin{prf}
The proof is briefly sketched in Section \ref{se:large}. The details, however, are left for the extended version of the paper.
\end{prf}

\subsection{Asymptotic Hardening Property}
Theorem \ref{thm:1} gives a simple approximation for several performance measures of the system in the large system limit. In fact, one can employ $\mai_{\mathrm{asy}}$ to approximate the ergodic capacity or outage probability of the system when $L$ is large enough. In addition to these applications, our main result depicts the asymptotic hardening property of the proposed setup. To illustrate the property, let $L$ grow large. In this case, $\xi_L$ grows proportional to $\log L$, and thus, $\beta_L$ tends to $1$. Consequently, $\mu_L$ reads
\begin{align}
\mu_L = \mao\left(\log{\log L}\right)
\end{align}
and $\sigma_L$ converges to zero. This observation indicates that $\mai$ asymptotically converges to a deterministic variable growing large proportional to $\log {\log L}$, and therefore, concludes the asymptotic hardening property of the system. More precisely, it states that, for large displacement supports, the observed channel between the transmit array and receive antenna, under the positioning protocol considered in Section \ref{ssec:local}, converges to a deterministic channel whose expected performance measure increases with respect to $L$.

The asymptotic hardening property in this setup is observed even by employing a few number~of~transmit antennas. This is due to the fact that spatially reconfigurable antenna arrays with large displacement support virtually implement massive \ac{mimo} systems in which a finite number of transmit antennas have been selected.\vspace*{1mm}

\section{Numerical Results}
\label{sec:application}
\begin{figure}[t]
\hspace*{-1.2cm}  
\centering
\resizebox{1.12\linewidth}{!}{
\pstool[width=.35\linewidth]{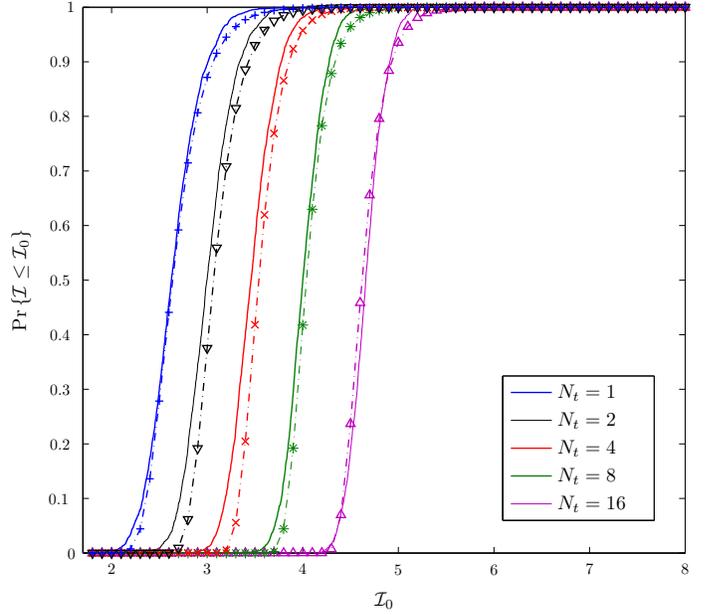}{

\psfrag{Nt=10AAA}[l][l][0.25]{$N_t=1$}
\psfrag{Nt=20AAA}[l][l][0.25]{$N_t=2$}
\psfrag{Nt=40AAA}[l][l][0.25]{$N_t=4$}
\psfrag{Nt=80AAA}[l][l][0.25]{$N_t=8$}
\psfrag{Nt=16AAA}[l][l][0.25]{$N_t=16$}

\psfrag{0}[r][c][0.2]{$0$}
\psfrag{0.1}[r][c][0.2]{$0.1$}
\psfrag{0.2}[r][c][0.2]{$0.2$}
\psfrag{0.3}[r][c][0.2]{$0.3$}
\psfrag{0.4}[r][c][0.2]{$0.4$}
\psfrag{0.5}[r][c][0.2]{$0.5$}
\psfrag{0.6}[r][c][0.2]{$0.6$}
\psfrag{0.7}[r][c][0.2]{$0.7$}
\psfrag{0.8}[r][c][0.2]{$0.8$}
\psfrag{0.9}[r][c][0.2]{$0.9$}
\psfrag{1}[r][c][0.2]{$1$}
%
%%x-axis
\psfrag{i}[c][b][0.25]{$\Pr\set{\mai \leq \mai_0}$}
\psfrag{Pr}[c][b][0.25]{${ \mai_0}$}
\psfrag{2}[c][b][0.2]{$2$}
\psfrag{3}[c][b][0.2]{$3$}
\psfrag{4}[c][b][0.2]{$4$}
\psfrag{5}[c][b][0.2]{$5$}
\psfrag{6}[c][b][0.2]{$6$}
\psfrag{7}[c][b][0.2]{$7$}
\psfrag{8}[c][b][0.2]{$8$}
\psfrag{2.6}[c][b][0.2]{$2.6$}
\psfrag{2.8}[c][b][0.2]{$2.8$}
}}
\caption{Comparison of empirical \ac{cdf} of $\mai$ and its Gumbel approximation given by Theorem~\ref{thm:1} for various number of transmit antennas. The solid and dashed lines~indicate the empirical and approximated \ac{cdf}, respectively. \ac{snr} is set to be $0$~dB, and $L=128$.}
\label{fig:2}
\end{figure}
%As it has been mentioned, Theorem~\ref{thm:1} gives a good approximation of the system performance in the large system limit.
In order to investigate the accuracy of the main result,~we plot the \ac{cdf} of $\mai_{\mathrm{asy}}$, and compare it with the empirical \ac{cdf} of $\mai$ evaluated numerically. Fig.~\ref{fig:2} demonstrates the empirical \ac{cdf}s of $\mai$ and the corresponding approximations for various numbers of transmit antennas when $L=128$. The empirical distributions are denoted by the solid lines and obtained via $20000$ channel realizations; moreover, \ac{snr} is set to be $0$~dB. As Fig.~\ref{fig:2} depicts, the type-one Gumbel distribution given by Theorem~\ref{thm:1} accurately approximates the empirical distribution of the mutual information. The figure, moreover, shows a jump in the probability from zero to one within~a~short~interval of $\mai_0$ which indicates the hardening property. %For the special case of $N_t=1$, the figure meets the results of \cite{bai2007channel}.

\begin{figure}[t]
\hspace*{-1.2cm}  
\centering
\resizebox{1.12\linewidth}{!}{
\pstool[width=.35\linewidth]{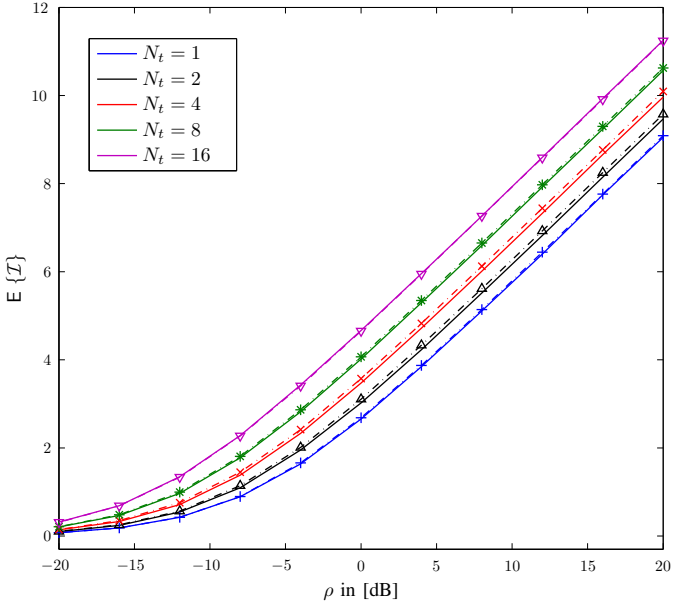}{

\psfrag{Nt=10AAA}[l][l][0.25]{$N_t=1$}
\psfrag{Nt=20AAA}[l][l][0.25]{$N_t=2$}
\psfrag{Nt=40AAA}[l][l][0.25]{$N_t=4$}
\psfrag{Nt=80AAA}[l][l][0.25]{$N_t=8$}
\psfrag{Nt=16AAA}[l][l][0.25]{$N_t=16$}
%y-axis

\psfrag{3}[r][c][0.2]{$0$}
\psfrag{2}[r][c][0.2]{$2$}
\psfrag{4}[r][c][0.2]{$4$}
\psfrag{6}[r][c][0.2]{$6$}
\psfrag{8}[r][c][0.2]{$8$}
\psfrag{11}[r][c][0.2]{$10$}
\psfrag{12}[r][c][0.2]{$12$}
%
%%x-axis
\psfrag{E}[c][b][0.25]{$\E \set{\mai}$}
\psfrag{rho}[c][c][0.25]{$\rho$ in [dB]}
\psfrag{-20}[c][b][0.2]{$-20$}
\psfrag{-15}[c][b][0.2]{$-15$}
\psfrag{-10}[c][b][0.2]{$-10$}
\psfrag{0}[c][b][0.2]{$0$}
\psfrag{5}[c][b][0.2]{$5$}
\psfrag{-5}[c][b][0.2]{$-5$}
\psfrag{10}[c][b][0.2]{$10$}
\psfrag{15}[c][b][0.2]{$15$}
\psfrag{20}[c][b][0.2]{$20$}

}}
\caption{Ergodic capacity versus \ac{snr}. The solid lines are plotted via numerical simulations, and the dashed lines denote the Gumbel approximations of the ergodic capacity given bu Theorem \ref{thm:1}. The size of the virtual channel is set to be $L=128$.}
\label{fig:3}
\end{figure}
As we mentioned in Section \ref{se:result}, for cases in which the \ac{csi} is only available at the receiver side, the input-output mutual information determines the maximum achievable transmit rate over the observed channel. In this case, the ergodic capacity represents the asymptotic average capacity of the channel and is defined as the expectation of $\mai$. In Fig.~\ref{fig:3}, the ergodic capacity is plotted in terms of the average \ac{snr} per antenna for several numbers of transmit antennas assuming the virtual channel to be of the size $L=128$. As the figure shows, the approximation given by Theorem \ref{thm:1} tracks the simulation results with less than $1.2\%$ deviation for a rather large range of \ac{snr}s.
\section{Sketch of the Proof}
\label{se:large}
%We assume the antennas transmit \ac{iid} Gaussian symbols with equal power. In this case, due to the input average power constraint, $\bx\sim\mathcal{CN} (\boldsymbol{0}, N_t^{-1} \mI_{N_t})$. The input-output mutual information, therefore, reads
%\begin{align}
%\mai \coloneqq \mathrm{I} \left( \bx;\by|\mH(\setS) \right)= \log \left( 1 + \frac{\rho}{N_t} \norm{\mH_{m^*}}^2 \right). \label{eq:mai}
%\end{align}
%In case that the \ac{csi} is available only at the receiver side, the input-output mutual information in \eqref{eq:mai} represents the maximum achievable rate over the observed channel $\tmH$. However, when the \ac{csi} is available at the both sides, $\mai$ determines only an achievable rate which is a lower bound on the capacity.
%
%The mutual information in \eqref{eq:mai} is in general a random variable whose distribution depends on the distribution of $\mH_{m^*}$.
In this section, we briefly illustrate the proof of Theorem~\ref{thm:1}. The details of derivations, however, are skipped here and left for the extended versions of the paper.

To start with the large-system analysis, let us define the sequence of random variables $\set{z_m}$ such that $z_m\coloneqq \mH_m \mH_m^\her$ and denote its maximum by $z_\m$, i.e., $z_{\m} \coloneqq \max_m z_m$ for $m\in[1:L-N_t+1]$. The random variable $z_\m$ represents $\norm{\mH_{m^*}}^2$ in \eqref{eq:mai}, and therefore, its distribution determines the distribution of $\mai$. Consequently, our problem reduces to the problem of evaluating the extreme value distribution of a sequence of random variables. For \ac{iid} sequences, the extreme value distribution is given by Fisher-Tippet law \cite{arnold1992first}. However, considering our problem, $\set{z_m}$ is an $N_t$-dependent sequence of random variables with identically distributed entries. To determine the extreme value distribution of $\set{z_m}$, we invoke the result of \cite{watson}, and show that Fisher-Tippet law also holds for our case.
\subsection{Fisher-Tippet Law}
Consider an \ac{iid} sequence $\set{x_m}$ whose entries are distributed with the \ac{cdf} $\rmF_x$. For this sequence, let us denote the extreme value by $x_\m$. The \ac{cdf} of $x_\m$, therefore, reads
\begin{subequations}
\begin{align}
\rmF_{x_\m}(x)&=\Pr \set{x_\m<x }\\
&=\Pr \set{ x_1<x, \ldots , x_m<x } =\left[\rmF_x(x)\right]^m.
\end{align}
\end{subequations}
The above derivation determines the extreme value distribution; however, it does not enables us to determine the asymptotic limit, i.e., the limit as $m\uparrow\infty$, explicitly. Fisher-Tippet law states that $x_\m$, as $m$ grows large, converges to the large limit of a random sequence in distribution which is a deterministic function of a random variable with the Fisher-Tippet distribution. To illustrate the law, let $m$ tends to infinity, and assume that the deterministic sequences $\set{c_m}$ and $\set{d_m}$ exist, such that
\begin{align}
\lim_{m\uparrow\infty} \left[ \rmF_x ( c_m x + d_m ) \right]^m = \rmF_h(x)
\end{align}
for some \ac{cdf} $\rmF_h$. In this case, one can conclude that
\begin{align}
\frac{x_\m-d_m}{c_m} \stackrel{\mathrm{d}}{\longrightarrow} h
\end{align}
as $m \uparrow \infty$, where $\stackrel{\mathrm{d}}{\longrightarrow}$ indicates convergence in distribution, and $h$ is a random variable distributed with \ac{cdf} $\rmF_h$. Fisher-Tippet law indicates that the above convergence argument holds for any given \ac{cdf} $\rmF_x$ in which
\begin{enumerate}[label=(\alph*)]
\item the normalization sequences $\set{c_m}$ and $\set{d_m}$ exist such that ${c_m \in \setR^+ }$ and ${d_m\in\setR}$, and
\item the converging \ac{cdf} $\rmF_h$ is one of the type-one Gumbel, Fr{\'e}chet, or Weibull \ac{cdf}s.
\end{enumerate}
%\begin{definition}[Domain of Maximal Attraction]
%For the given \ac{cdf} $\rmF_h$, the domain of maximal attraction is defined as the set of all \ac{cdf}s $\rmF_x$ in which at any continuity point of $\rmF_h$
%for some deterministic sequences $\set{c_m \in \setR^+ }$ and $\set{d_m\in\setR}$.
%\end{definition}

Regarding the sequence $\set{z_m}$, the entries are identically distributed, but not independent. Therefore, the above extreme value convergence arguments for this case need to be further discussed. In the sequel, we show that the extreme value of $\set{z_m}$ follows Fisher-Tippet law with the converging type-one Gumbel distribution.

\subsection{Asymptotic Extreme Value Distribution of $\set{z_m}$}
The sequence $\set{z_m}$ is an $N_t$-dependent sequence with identically distributed entries. As the entries of $\mH_m$ are \ac{iid} zero-mean unit-variance complex Gaussians, $z_m$, for any $m$, is a chi-squared random variable with $2 N_t$ degrees of freedom, i.e., $z_m \sim \rmF_z$ where
\begin{align}
\rmF_z(z)=\int_0^z \frac{w^{N_t-1} e^{-w}}{(N_t-1)!} \hspace{.3mm} \dif w.
\end{align}
It is therefore straightforward to conclude that $\set{z_m}$ is an stationary process. In order to study the extreme value distribution of $\set{z_m}$, one needs to extend the scope of Fisher-Tippet law to a larger set of sequences. To do so, define the dependency measure
\begin{align}
\Delta_{x}(k;u)\coloneqq \dfrac{\Pr \set{ \min\set{x_m, x_{m+k}} > u}}{\Pr\set{x_m > u}} \label{eq:Delta}
\end{align}
for some stationary sequence $\set{x_m}$, where the index $m$ in $\Delta_{x}(k;u)$ is dropped, due to the fact that the joint statistics  of stationary sequences only depend on the difference of the indices. For \ac{iid} sequences, $\Delta_{x}(\cdot;\cdot)$ reads
\begin{align}
\Delta_{x}(k;u)= \Pr\set{x_m > u} = 1-\rmF_x(u)
\end{align}
which tends to zero, as $u$ grows large. Therefore, one can write
\begin{align}
\lim_{u\uparrow\infty} \max_{k\neq 0} \Delta_x(k;u)=0 \label{eq:limit}
\end{align}
for $k\in\setZ$. In general, it is shown that the scope of Fisher-Tippet law extends to the set of $\sfM$-dependent stationary~random~sequences which satisfy the constraint in \eqref{eq:limit}; see \cite{watson}.

For $\set{z_m}$, the symbols $z_m$ and $z_{m+k}$ are independent as $k\geq N_t$, and $\Delta_z(k;u)$ converges to zero for these values of $k$, as $u$ grows large. Thus, the constraint in \eqref{eq:limit} reduces to
\begin{align}
\lim_{u\uparrow\infty} \max_{0<k<N_t} \Delta_z(k;u)=0. \label{eq:21}
\end{align}
Considering the definition of the sequence $\set{z_m}$, the constraint in \eqref{eq:21} is shown to hold, and therefore, the distribution of $z_{\mathsf{max}}$ is determined via Fisher-Tippet law. At this point, we need to determine the converging law $\rmF_h$ for $\set{z_m}$, as well as the deterministic sequences $\set{c_m > 0}$ and $\set{d_m}$ which satisfy
\begin{align}
\lim_{m\uparrow\infty} \rmF_{z_\m} ( c_m z + d_m ) = \rmF_h(z)
\end{align}
By standard analytical arguments, it is shown that the converging law of a sequence of chi-squared random variables is $\mathrm{Gumbel}\hspace*{.3mm} [0;1]$. Consequently, $\set{c_m }$ and $\set{d_m}$ can be determined in terms of $\rmF_z$ as in Fisher-Tippet law. The obtained sequences, however, have a low rate of convergence, and thus, the asymptotic distribution cannot approximate the result well for a finite length of the sequence. In order to improve the convergence rate, we modify the sequences $\set{c_m }$ and $\set{d_m}$ by invoking the results given in \cite{gasull2013maxima}. After some trivial lines of derivations, $\set{d_m}$ is determined as
\begin{align}
\hspace*{-.7mm} d_m \hspace*{-.7mm} = \hspace*{-.7mm} a_m \hspace*{-.7mm} + \hspace*{-.7mm} \delta_{N_t,1} \frac{N_t \hspace*{-.7mm} - \hspace*{-.7mm} 1}{a_m} \left[ 1 \hspace*{-.7mm} + \hspace*{-.7mm} \left(N_t\hspace*{-.7mm} +\hspace*{-.7mm} a_m \hspace*{-.7mm} - \hspace*{-.7mm} 1 \right) \loge \set{\frac{a_m}{N_t\hspace*{-.7mm} -\hspace*{-.7mm} 1}} \right]
\end{align}
where $\delta_{N_t,1}$ returns zero at $N_t=1$ and one elsewhere , and the sequence $\set{a_m}$ is defined as
\begin{align}
a_m \coloneqq \loge \set{\frac{m}{(N_t-1)!}} +\delta_{N_t,1} (N_t-1) \hspace*{.5mm} \loge(N_t-1),
\end{align}
and the sequence $\set{c_m}$ is given by
\begin{align}
c_m = \frac{ d_m^{\hspace*{.2mm}2}+ \left( N_t-1 \right) d_m}{d_m^{\hspace*{.2mm}2}-(N_t-1)(N_t-2)}.
\end{align}
Therefore, one can conclude that
\begin{align}
\frac{z_\m - d_m}{c_m} \stackrel{\mathrm{d.}}{\longrightarrow} z_0
\end{align}
where $z_0 \sim e^{-e^{-z_0}}$. Substituting in \eqref{eq:mai}, as $L$ grows large, the input-output mutual information converges in distribution~to
\begin{align}
\mai_{\mathrm{asy}} \coloneqq \log \set{ 1+\rho \left( c_{L-N_t+1} z_0 + d_{L-N_t+1} \right) }.
\end{align}
By defining $\xi_L \coloneqq a_{L-N_t+1}$ and $\phi_L \coloneqq d_{L-N_t+1}-a_{L-N_t+1}$, and the non-negative coefficient $\beta_L \coloneqq c_{L-N_t+1}$, the asymptotic input-output mutual information is written as
\begin{subequations}
\begin{align}
\mai_{\mathrm{asy}} & = \log \set{ 1+\rho \left( \beta_L \gamma + \xi_L + \phi_L \right) + \rho \tilde{z}_0 }\\
&= \log \set{ 1+\rho \left( \beta_L \gamma + \xi_L + \phi_L \right) } + \log \set{1 + \hat{z}_0 } \label{eq:28a}
\end{align}
\end{subequations}
where $\gamma$ is Euler's constant, $\tilde{z}_0 \coloneqq z_0-\gamma$ is a zero-mean~type- one Gumbel random variable, and $\hat{z}_0$ is defined as
\begin{align}
\hat{z} \coloneqq \frac{\rho \tilde{z}_0}{1+\rho \left( \beta_L \gamma + \xi_L + \phi_L \right)}.
\end{align}
The second expression in the right hand side of \eqref{eq:28a}~is~further simply expanded as
\begin{align}
\log \set{1 + \hat{z}_0 } =\hat{z}_0 + \mao (\hat{z}_0^2). \label{eq:30}
\end{align}
The random variable $\hat{z}_0$ is a zero-mean random variable whose variance reduces with respect to $\loge^{-1} L$ as $L$ grows. Therefore, the second term in the right hand side of \eqref{eq:30} is a random term vanishing to zero with respect to $\loge^{-2} L$. Substituting \eqref{eq:30} in \eqref{eq:28a}, Theorem \ref{thm:1} is finally concluded.

%\section{Conclusion}
%\label{conclusion}
%Conclusion

\bibliography{ref}

\begin{thebibliography}{10}
\providecommand{\url}[1]{#1}
\csname url@samestyle\endcsname
\providecommand{\newblock}{\relax}
\providecommand{\bibinfo}[2]{#2}
\providecommand{\BIBentrySTDinterwordspacing}{\spaceskip=0pt\relax}
\providecommand{\BIBentryALTinterwordstretchfactor}{4}
\providecommand{\BIBentryALTinterwordspacing}{\spaceskip=\fontdimen2\font plus
\BIBentryALTinterwordstretchfactor\fontdimen3\font minus
  \fontdimen4\font\relax}
\providecommand{\BIBforeignlanguage}[2]{{%
\expandafter\ifx\csname l@#1\endcsname\relax
\typeout{** WARNING: IEEEtran.bst: No hyphenation pattern has been}%
\typeout{** loaded for the language `#1'. Using the pattern for}%
\typeout{** the default language instead.}%
\else
\language=\csname l@#1\endcsname
\fi
#2}}
\providecommand{\BIBdecl}{\relax}
\BIBdecl

\bibitem{rappaport2013millimeter}
T.~S. Rappaport, S.~Sun, R.~Mayzus, H.~Zhao, Y.~Azar, K.~Wang, G.~N. Wong,
  J.~K. Schulz, M.~Samimi, and F.~Gutierrez, ``Millimeter wave mobile
  communications for 5G cellular: It will work!'' \emph{IEEE access}, vol.~1,
  pp. 335--349, 2013.

\bibitem{marzetta2010noncooperative}
T.~L. Marzetta, ``Noncooperative cellular wireless with unlimited numbers of
  base station antennas,'' \emph{IEEE Transactions on Wireless Communications},
  vol.~9, no.~11, pp. 3590--3600, 2010.

\bibitem{molisch2005capacity}
A.~F. Molisch, M.~Z. Win, Y.-S. Choi, and J.~H. Winters, ``Capacity of MIMO
  systems with antenna selection,'' \emph{IEEE Transactions on Wireless
  Communications}, vol.~4, no.~4, pp. 1759--1772, 2005.

\bibitem{li2014energy}
H.~Li, L.~Song, and M.~Debbah, ``Energy efficiency of large-scale multiple
  antenna systems with transmit antenna selection,'' \emph{IEEE Transactions on
  Communications}, vol.~62, no.~2, pp. 638--647, 2014.

\bibitem{asaad2017tas}
S.~Asaad, A.~Bereyhi, R.~R. M{\"u}ller, and A.~M. Rabiei, ``Asymptotics of
  transmit antenna selection: Impact of multiple receive antennas,''
  \emph{International Conference on Communications (ICC)}, 2017.

\bibitem{gheethan2015passive}
A.~A. Gheethan, A.~Dey, and G.~Mumcu, ``Passive feed network designs for
  microfluidic beam-scanning focal plane arrays and their performance
  evaluation,'' \emph{IEEE Transactions on Antennas and Propagation}, vol.~63,
  no.~8, pp. 3452--3464, 2015.

\bibitem{dey2016microfluidically}
A.~Dey and G.~Mumcu, ``Microfluidically controlled frequency-tunable monopole
  antenna for high-power applications,'' \emph{IEEE Antennas and Wireless
  Propagation Letters}, vol.~15, pp. 226--229, 2016.

\bibitem{palomo2016microfluidically}
T.~Palomo and G.~Mumcu, ``Microfluidically reconfigurable metallized plate
  loaded frequency-agile RF bandpass filters,'' \emph{IEEE Transactions on
  Microwave Theory and Techniques}, vol.~64, no.~1, pp. 158--165, 2016.

\bibitem{yilmaz2016millimeter}
M.~Yilmaz, E.~Guvenkaya, G.~Mumcu, and H.~Arslan, ``Millimeter-wave wireless
  channel control using spatially adaptive antenna arrays,'' \emph{IEEE
  Communications Letters}, 2016.

\bibitem{hochwald2004multiple}
B.~M. Hochwald, T.~L. Marzetta, and V.~Tarokh, ``Multiple-antenna channel
  hardening and its implications for rate feedback and scheduling,'' \emph{IEEE
  transactions on Information Theory}, vol.~50, no.~9, pp. 1893--1909, 2004.

\bibitem{bai2007channel}
D.~Bai, P.~Mitran, S.~S. Ghassemzadeh, R.~R. Miller, and V.~Tarokh, ``Rate of
  channel hardening of antenna selection diversity schemes and its implication
  on scheduling,'' \emph{IEEE Transactions on Information Theory}, vol.~55,
  no.~10, pp. 4353--4365, 2009.

\bibitem{arnold1992first}
B.~C. Arnold, N.~Balakrishnan, and H.~N. Nagaraja, \emph{A first course in
  order statistics}.\hskip 1em plus 0.5em minus 0.4em\relax Siam, 1992,
  vol.~54.

\bibitem{watson}
G.~Watson, ``Extreme values in samples from m-dependent stationary stochastic
  processes,'' \emph{The Annals of Mathematical Statistics}, pp. 798--800,
  1954.

\bibitem{gasull2013maxima}
A.~Gasull, J.~A. L{\'o}pez-Salcedo, and F.~Utzet, ``Maxima of Weibull-like
  distributions and the Lambert W function,'' \emph{arXiv preprint~arXiv: 1308.5534}, 2013.

\end{thebibliography}
\bibliographystyle{IEEEtran}

\begin{acronym}
\acro{mmw}[mmWave]{millimeter Wave}
\acro{mimo}[MIMO]{Multiple-Input Multiple-Output}
\acro{csi}[CSI]{Channel State Information}
\acro{awgn}[AWGN]{Additive White Gaussian Noise}
\acro{iid}[i.i.d.]{independent and identically distributed}
\acro{ut}[UT]{User Terminal}
\acro{bs}[BS]{Base Station}
\acro{tas}[TAS]{Transmit Antenna Selection}
\acro{lse}[LSE]{Least Squared Error}
\acro{rhs}[r.h.s.]{right hand side}
\acro{lhs}[l.h.s.]{left hand side}
\acro{wrt}[w.r.t.]{with respect to}
\acro{rs}[RS]{Replica Symmetry}
\acro{rsb}[RSB]{Replica Symmetry Breaking}
\acro{papr}[PAPR]{Peak-to-Average Power Ratio}
\acro{rzf}[RZF]{Regularized Zero Forcing}
\acro{snr}[SNR]{Signal-to-Noise Ratio}
\acro{rf}[RF]{Radio Frequency}
\acro{mf}[MF]{Match Filtering}
\acro{pdf}[PDF]{Probability Density Function}
\acro{cdf}[CDF]{Cumulative Distribution Function}
\end{acronym}

\end{document}